\documentclass{aa}
\usepackage{epsfig}

\def\ltsima{$\; \buildrel < \over \sim \;$}
\def\simlt{\lower.5ex\hbox{\ltsima}}
\def\gtsima{$\; \buildrel > \over \sim \;$}
\def\simgt{\lower.5ex\hbox{\gtsima}}

\begin{document}
   \thesaurus{03 (13.25.2;11.19.1;12.04.2)}
  \title{Beyond the standard model for the cosmic X--ray background}

   \author{R. Gilli
                \inst{1}
   \and    G. Risaliti
		\inst{1}		
   \and    M. Salvati
              \inst{2}
}
   \offprints{R. Gilli}

  \institute {Dipartimento di Astronomia e Scienza dello Spazio, Universit\`a 
di Firenze, Largo E. Fermi 5, I--50125 Firenze, Italy,
gilli@arcetri.astro.it, risaliti@arcetri.astro.it
     \and   Osservatorio Astrofisico di Arcetri, Largo E. Fermi 5,
I--50125 Firenze, Italy, salvati@arcetri.astro.it
}

   \date{Received / Accepted }

\titlerunning{Beyond the standard model for the XRB}
\authorrunning{R. Gilli et al.}

   \maketitle

   \begin{abstract}

The synthesis model for the cosmic X--ray background (XRB) --based on
the integrated emission of Active Galactic Nuclei (AGNs)-- is 
complemented with new observational results. We adopt the most recent 
estimates of the AGN X--ray luminosity function and evolution. We adopt the 
column density distribution of type 2 AGNs observed in the local Universe, 
instead of describing it as a set of free parameters. We maintain the standard 
assumptions that type 2 AGNs have the same luminosity function --apart from
a constant factor-- and the same evolution of type 1s, and that the spectral
shapes of both types are independent of redshift.
We explore various parametrizations of the data,
and in all cases we find that the XRB can be fitted, but the number density 
ratio of type 2s to type 1s must be higher than the local value, and/or the 
hard X--ray counts in the 2--10 keV band and the preliminary BeppoSAX counts 
in the 5--10 keV band are underestimated at a level of a few $\sigma$.
These results consistently suggest that type 2 AGNs could undergo a faster 
evolution than type 1s, or other sources with hard X--ray spectra might 
contribute to the XRB at intermediate or high redshifts.

   \end{abstract}


%

\section{Introduction}

The cosmic X--ray background (XRB) above $\sim 1$ keV is the result of the
integrated emission of discrete sources, since the contribution of any 
intergalactic hot medium must be negligible (Wright et al. 1994).
In the soft X--ray band from 0.5 to 2 keV the largest fraction of the XRB has 
already been resolved into sources (Hasinger et al. 1998), most of which turned
out to be broad line active nuclei (Schmidt et al. 1998), i.e. 
Quasi Stellar Objects (QSOs) and Seyfert galaxies of type~1.
The spectra of these sources are however too steep to reproduce also the hard
XRB at several tens of keV, where the bulk of the energy resides, and a
population of objects with flatter spectra is therefore required.  

The most popular synthesis models of the XRB are based on the so--called
unification schemes for Active Galactic Nuclei (AGNs), where the orientation 
of a molecular torus
surrounding the nucleus determines the classification of the source. 
At a zeroth--order approximation level, sources observed along lines of
sight free from the torus
obscuration should have unabsorbed X--ray spectra and optical
broad lines (type 1 AGNs), while sources seen through the torus should
have absorbed
X--ray spectra and appear as narrow line objects in
the optical (type 2 AGNs, e.g. Seyfert 2 galaxies).

In this framework type 2 AGNs provide a natural class of sources
with X--ray spectra flattened by absorption. The intrinsic
X--ray luminosity function (XLF) of type 2 objects is unknown and has been
usually assumed to be the same as the one derived for type 1s (e.g. Boyle et
al. 1993), apart from a normalization factor. The cosmological
evolution has also been taken identical for type 1s and type 2s.   
Under these assumptions it has been shown that the broad band 3--100 keV
spectrum of the XRB can be reproduced by an appropriate mix of 
unabsorbed and absorbed AGNs (Matt \& Fabian 1994; Madau, Ghisellini \&
Fabian 1994; Comastri et al. 1995, hereafter Co95). The number ratio $R$
of type 2 to type 1 objects, as well as the distribution of the
absorbing column densities $N_{\rm H}$, are key parameters of the models;
these have been assumed to be independent of redshift and of intrinsic
source 
luminosity, and have been treated as free parameters in the fitting
procedure.
Since the overall parameter space of the models is quite large and a good
fit to the XRB can be obtained with different set of values, it is
important to compare the model predictions with the largest number of
observational constraints. Indeed, Co95
showed that the source counts in the 0.5--2 keV
and 2--10 keV energy bands, as well as the redshift distributions, could
successfully be reproduced by their model.     

Very recently an additional set of observational constraints
has become available.
Deep surveys from ROSAT have extended our knowledge to 
the low luminosity part of the AGN XLF (Miyaji, Hasinger \& Schmidt
1999a, hereafter Mi99a).
Contrary to previous results (Boyle et al. 1993; Page et al. 1996; Jones 
et al. 1997) a pure luminosity evolution (PLE) of AGNs with redshift is no 
longer consistent with the data, and a luminosity
dependent density evolution (LDDE) is required. 
From the X--ray data of an optically selected sample of Seyfert galaxies
Risaliti, Maiolino \& Salvati (1999) have determined the $N_{\rm H}$
distribution for local
Seyfert 2 galaxies, pointing out that a significant fraction of sources
have columns exceeding $N_{\rm H}=10^{25}$ cm$^{-2}$ and are therefore
completely thick to Compton scattering. The $R$ ratio between type 2s and
type 1s has been determined in the local Universe for low luminosity AGNs, 
i.e. Seyfert galaxies (Maiolino \& Rieke 1995), while  
the existence of a relevant number of high luminosity absorbed
sources, the so--called QSO~2s, which is a basic assumption of
previous models, is still uncertain (Akiyama et al. 1998).
An observational constraint to
the QSO~2 number density can be obtained from the infrared source
counts. Indeed, QSO~2s are expected to
have strong infrared counterparts, since
the dust present in the torus should re--emit in the IR band the
nuclear radiation absorbed by the gas.
The ultraluminous infrared galaxies (ULIRGs) discovered by IRAS are the 
only local objects with QSO--like bolometric luminosities (Soifer et al. 
1986; Kim \& Sanders 1998).
Thus, even if all ULIRGs were powered by a hidden AGN, 
the local QSO~2s could not be more numerous than ULIRGs. 
Finally, source counts in the 5--10 keV band have been derived for the
first time by the BeppoSAX satellite with the HELLAS survey 
(Giommi et al. 1998; Comastri et al. 1999).

In the present paper we test the standard synthesis 
model to verify if it remains compatible with the new data. These data
leave still some latitude to important parameters of the model, and various
choices are possible to fit the XRB equally well. However, in all cases
we find moderate but consistent evidence that at least some of the standard 
assumptions have to be relaxed: extra hard spectrum AGNs are needed at
intermediate or high redshifts, in addition to those expected in the usual
scenario. The additional sources could be analogous to
local Seyfert 2s, if they evolve faster than type 1s, or they could be
other astrophysical sources not yet enlisted among the contributors to the
XRB. We discuss the observations which could distinguish between
the alternatives.
  
Throughout this paper the deceleration parameter and the Hubble constant 
are given the values $q_{0}=0.5$ and $H_{0}=50$ km s$^{-1}$ Mpc$^{-1}$.

\section{AGN X--ray properties}

\subsection{The spectra}

After the observations of X--ray satellites like GINGA, ASCA and
BeppoSAX, different components 
have been recognized in the X--ray spectra of
AGNs. Starting with Sey 1 galaxies, the basic component is a power law
with energy
spectral index $\alpha\sim0.9$ (Nandra et al. 1997a) and an
exponential  
cut off at high energies. A mean value for the $e$--folding energy can
probably be set at $\sim 300$ keV, although the observed dispersion is
very high (Matt 1998).   
Some of the primary radiation is reprocessed by an accretion disc and/or  
the torus around the nucleus, producing a flattening of the spectral slope 
above $\sim 10$ keV, and a strong iron line at 6.4 keV (Nandra \& Pounds 1994).  
Below 1--1.5 keV a radiation excess with respect to the power law
emission is detected in a large fraction of Sey 1s (sometimes resulting
from a misfit of the ``warm absorber'' component).     

The spectrum of QSO~1s is similar to that of Sey 1s, but there is no evidence
for the iron line and the reflection hump to be as common (Lawson \& Turner
1997). Assuming that the accretion disc produces most of the line and hump, 
Nandra et al. (1997b) ascribe these differences to a higher ionization state 
of the disc in higher accretion rate sources, so that in QSO~1s the 
spectral features due to photoelectric processes are quenched. Recently, 
Vignali et al. (1999) have derived a mean spectral slope of $\langle
\alpha \rangle =0.67\pm 0.11$ from a sample of 5 QSO~1s at redshifts above 
2. Although the statistics is poor, this result seems to suggest that the
spectra of high redshift QSOs are flatter than those of local ones.

In Sey 2 galaxies the power law is cut off by photoelectric absorption at
energies increasing with the column density of the intercepted torus. 
For highly absorbed objects the X--ray luminosity may be dominated by that 
fraction of the nuclear radiation which is reflected off the torus surface 
towards the observer. When $N_{\rm H}>10^{25}$ cm$^{-2}$ the
obscuring medium is completely thick to Compton scattering 
and the spectrum is a pure reflection continuum as described by Lightman
\& White (1988), with a 2--10 keV luminosity about two orders of
magnitude lower than that of Sey 1s (Maiolino et al. 1998). On the
contrary, when $N_{\rm H}<10^{24}$ cm$^{-2}$ the medium is Compton--thin
and the spectrum is dominated by the component transmitted through the
torus. In the range
$10^{24}<N_{\rm H}<10^{25}$ cm$^{-2}$ both a transmitted and a reflected
component contribute to the observed luminosity, the Circinus galaxy
(Matt et al. 1999) being a typical example.          
Also Sey 2 galaxies often have soft emission in excess of
the absorbed power law (Turner et al. 1997).
These soft excesses are however two orders of magnitude weaker than those
of Sey 1s of the same intrinsic luminosity and their nature is still
unclear (probably scattered or starburst radiation). 

\subsection{The XLF and cosmological evolution}

The most recent results about the AGN XLF and cosmological evolution
have been obtained by Mi99a by combining data from several
ROSAT surveys. Down to a limiting flux of $10^{-15}$ erg s$^{-1}$
cm$^{-2}$, reached by the deep survey in the Lockman Hole, they
collected a sample of about 670 sources,
which is the largest X--ray selected sample of AGNs presently available. 
The local XLF is described with a smoothed double power law of the
following form:

\begin{displaymath}
\phi(L_{\rm x})=\frac{{\rm d}\,\Phi(L_{\rm x})}{{\rm
d\,log}L_{\rm x}}={A}\,\left[(L_{\rm x}/{L_*})
      ^{{\gamma_1}}
         +(L_{\rm x}/{L_*})^{{\gamma_2}}
        \right]^{-1}, 
\end{displaymath}

\noindent where $L_{\rm x}$ is the observed 0.5--2 keV X--ray luminosity, 
ranging from $10^{41.7}$ to $10^{47}$ erg s$^{-1}$.
The best fit values for the cosmology adopted here are: 
$A=(1.57 \pm 0.11)\times 10^{-6}$ Mpc$^{-3}$, 
$L_*=0.57^{+0.33}_{-0.19} \times 10^{44}$ erg s$^{-1}$, $\gamma_1=0.68\pm
0.18$ and $\gamma_2=2.26\pm 0.95$. 

The XLF has been found to evolve from redshift 0 up to $z_{cut}=1.51\pm
0.15$, with an evolution rate which drops at low
luminosities according to the factor:
\begin{eqnarray}
e(z,L_{\rm x})=        \left\{ 
        \begin{array}{ll}
        (1+z)^{\max(0,{p1}-{\alpha}({\rm log}\; 
          {L_{\rm a}} - {\rm log}\;L_{\rm x}))} 
             & L_{\rm x}<L_{\rm a} \\ 
        (1+z)^{{p1}} 
             & L_{\rm x}\ge L_{\rm a}\;;\\ 
        \end{array}
	\right.
       \nonumber
\end{eqnarray}

\noindent here $p1=5.4\pm 0.4$, $\alpha = 2.3\pm 0.8$ and 
${\rm log} L_{\rm a}=44.2$ (fixed).
The X--ray AGNs have been observed at redshifts up to $z=4.6$ and there is no
evidence for a decline in their space density beyond $z\sim 3$, unlike what
is found in optical (Schmidt, Schneider \& Gunn 1995) and radio surveys 
(Shaver et al. 1997). 

We note that the XLF parametrization of Mi99a 
is a preliminary result and is not a unique
representation of the ROSAT data. The extrapolation of the high
redshift XLF into the low luminosity range, where few data are available,
is not well constrained. Indeed, the number of
low luminosity, high redshift AGNs could be higher than 
the Mi99a representation (Hasinger et al. 1999). Another cause of
uncertainty is the possible presence of type 2 AGNs in the 
Mi99a sample: unlike previous works, where only (optical) type 1 AGNs 
were included, Mi99a do not discriminate between type 1s and type 2s; 
some of the latter could then appear in the ROSAT bandpass because of their 
soft excesses and, for sources at high redshifts, because of the
$K$--correction. Objects with type 2 optical spectra are relatively
rare in the ROSAT sample (Hasinger et al. 1999). As for the X--ray 
spectral type, within any given model the raw counts can be corrected
for the contribution of the absorbed sources, and these can
be subtracted from the XLF: in the following we consider also this approach,
and investigate the robustness of our conclusions with respect to the
correction. In general, the correspondence between optical and X--ray
spectral classification is broadly verified in the local Universe, albeit
with some blurring (see Section 2.3); at high redshifts the question is 
still unsettled.
 
Previous works about the XLF of AGNs used a PLE model both in the soft 
X rays, by combining observations from ROSAT and Einstein (Boyle
et al. 1993, 1994; Jones et al. 1997), and in the hard X rays from ASCA
data (Boyle et al. 1998a). In the Mi99a data the fit with a PLE model is 
rejected with a high significance. A pure density evolution model is
marginally rejected, and LDDE models are preferred, even if several variants
are still being discussed.

\subsection{The number and column densities of local type 2 AGNs}          

In the local Universe 5--10\% of the galaxies show Seyfert activity
(Maiolino \& Rieke 1995; Ho, Filippenko \& Sargent 1997). From
a sample of $\sim 90$ nearby Seyfert galaxies limited in the B
magnitude of the host galaxy, Maiolino \& Rieke (1995) derived an estimate
for the local ratio $R$ of type 2 to type 1 Seyferts.
From our point of view Seyfert types 1.8, 1.9 and 2, which
have flat X--ray spectra due to absorption by cold gas, can be grouped
as type 2 objects, while types 1, 1.2 and 1.5, which have steep X--ray
spectra without significant cold absorption ($N_{\rm H}<10^{21}$
cm$^{-2}$), can be grouped as type 1s. Here it is noted that
the relation between Seyfert type and X--ray absorption is not univocal.
By observing with ROSAT the complete sample of Piccinotti et
al. (1982), Schartel et al. (1997) showed that
at least a fraction of type 1 AGNs suffer from X--ray
absorption by more than $N_{\rm H}=10^{21}$ cm$^{-2}$.
However this fraction (grouping Seyfert 1, 1.2 and 1.5) is only 20\%, and
on average their $N_{\rm H}$ does not exceed $10^{22}$ cm$^{-2}$. The
inclusion of some 
moderate--absorption type 1s should not change significantly our results.   
  
By considering Seyfert types 1.8, 1.9 and 2 as type 2s, and Seyfert types 1, 
1.2 and 1.5 as type 1s, Maiolino \& Rieke found $R$=4.0$\pm$0.9, in agreement 
with the results of
Osterbrock \& Martel (1993) and, more recently, Ho et al. (1997). 

From the Maiolino \& Rieke sample Risaliti et al. (1999)
have derived a distribution of X--ray column densities for Sey 2s.
The selection of the sample by means of optical narrow emission lines, 
rather than in the X--rays,
should avoid biases against X--ray absorbed sources. It
turned out that most of the sources are affected by strong absorption,
$\sim 75\%$ of the objects having $N_{\rm H}>10^{23}$ cm$^{-2}$. 
Furthermore, a significant fraction of sources ($>25\%$) are
absorbed by $N_{\rm H}>10^{25}$ cm$^{-2}$. 
Their results are shown in Fig.~1 and compared with the $N_{\rm H}$
distribution assumed by Co95. 

\begin{figure}
\epsfig{file=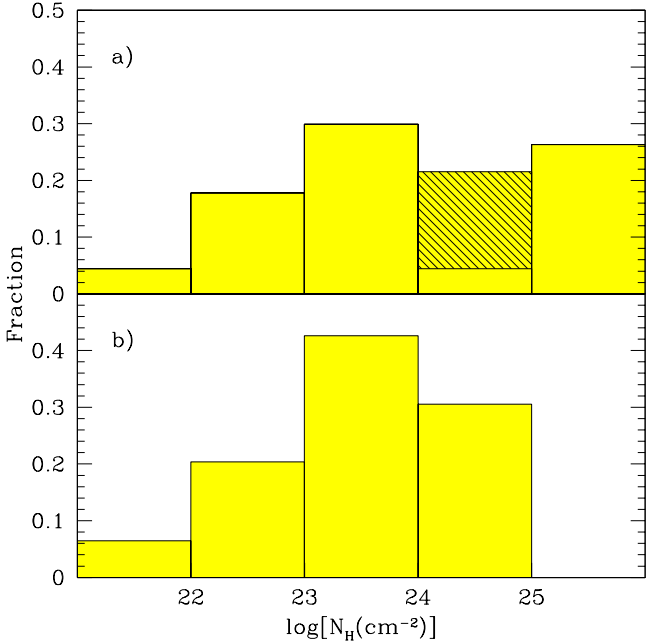,  width=9.cm, height=9.cm,
angle=0}
\caption{The normalized $N_{\rm H}$ distribution in Sey 2 galaxies
derived by Risaliti et al. (1999) [Panel a)], compared with
the distribution assumed by Co95 [Panel b)]. The shaded area represents
the fraction of sources for which only a lower limit of log$N_{\rm H}>24$
is available.}
\end{figure}

In the limited luminosity range of Seyfert galaxies Risaliti et al. (1999) 
did not find evidence of a correlation between absorption and luminosity.
In the wider luminosity domain which includes the QSOs the evidence 
is contradictory. Recent results from the IRAS 1--Jy survey (Kim \& Sanders
1998) show that the space density of ULIRGs at $z\stackrel{<}{_{\sim}}
0.1$ is similar to that of optically selected QSOs of
comparable bolometric luminosity. By considering that the number
of obscured  
QSOs cannot exceed that of ULIRGs, and assuming that every ULIRG is powered
only by nuclear activity, we can set a conservative upper limit of
2 to the $R$ ratio at high luminosities. 
The actual value of $R$ might be significantly lower. Indeed,
there is evidence that the fraction of AGN--powered ULIRGs 
decreases from 50\% for IR luminosities $L_{\rm
IR}\stackrel{>}{_{\sim}} 1.7\times 10^{46}$ erg s$^{-1}$ to 15\% below
this value (Lutz et al. 1998), the remaining ones being dominated by
starburst activity.

\section{The model} 

\subsection{The XRB spectrum}

Our models are completely analogous to those of the canonical lineage,
the only differences arising from updated input parameters. A key set 
of such parameters is the one referring to the XLF and its cosmological
evolution, and for model A1 we adopt the results of Mi99a.
Strictly speaking, the XLF of Mi99a refers to the observed 0.5--2 keV
luminosities and could be considered as the 0.5--2 keV XLF in the rest
frame only by assuming simple power law spectra with energy index
$\alpha=1$ (i.e. zero $K$--correction). Indeed, in the rest frame energy
range seen by ROSAT at different redshifts, the spectra of type
1 AGNs assumed in our models do not differ significantly from a
power law with $\alpha=1$. Since we refer the Mi99a XLF to unabsorbed
AGNs, without correcting for the contribution of absorbed AGNs to the
ROSAT counts, model A1 might be biased in favor of a soft XRB.
We discuss the strength of this bias in connection with models A2 and B 
in the following.

The absorption distribution in type 2 AGNs is no longer derived from best
fitting, instead it is taken equal to the local one, as measured by Risaliti 
et al. (1999). The objects for which only a lower
limit is available, $N_{\rm H}>10^{24}$ cm$^{-2}$, have been
assigned to the bin $10^{24}<N_{\rm H}<10^{25}$ cm$^{-2}$. 

Because of the evidence that the $R$ ratio decreases with the intrinsic 
luminosity of the AGNs, at least locally, we introduce a change with respect 
to the canonical scenario: the XLF is divided in two luminosity regions
as follows:

\begin{displaymath}
\phi(L_{\rm x})=\phi(L_{\rm x})e^{-\frac{L_{\rm x}}{L_s}}+\phi(L_{\rm
x})(1-e^{-\frac{L_{\rm x}}{L_s}})\;,
\end{displaymath}

\noindent with the 0.5--2 keV $e$--folding luminosity set 
equal to $L_s$=10$^{44.3}$ erg s$^{-1}$, following Miyaji, Hasinger \&
Schmidt (1999b; hereafter Mi99b). 
The first and second term represent the XLF of Sey 1s and QSO 1s,
respectively, and, apart from the exponential factors, are equal to
the Mi99a functions as given in Section 2.2. 
The XLF of Sey 2s and QSO 2s are $R_{\rm S}$ and $R_{\rm Q}$ times the XLF 
of the corresponding type~1 objects. In this parametrization we can explore
various hypotheses, including for instance the effects of eliminating
altogether the QSO 2s.

Following Co95, and the experimental evidence referred to in Section 2.1,
we assume that the basic spectrum for
type 1 AGNs is a power law with energy index $\alpha=0.9$ and 
exponential cut off with $e$--folding energy $E_c=320$ keV. Below 1.5 keV
the soft excess is modeled with a power law of index 1.3. A reflection
component from the accretion disc has been included for Sey 1s with
relative normalization $f_d=1.29$ (Co95). 
Beside the disc, we have also included for Sey 1s a
torus reflection component which is 
normalized in accordance with the prescriptions of
Ghisellini, Haardt \& Matt (1994). In type 1 AGNs the relative 
contribution of the torus at 30 keV is 29\% and 55\% for $N_{\rm H}=10^{24}$
and $10^{25}$ cm$^{-2}$, respectively. If we assume that the column density
of the torus is approximately the same for all obscured lines of sight,
from the measured $N_{\rm H}$ distribution we find that the torus contributes
on the average 28\% at 30 keV. The same disc and torus reflection
components of Sey 1s have been included also in the QSO spectra:
this is against the evidence at low redshifts, but mimics the harder
power law seen at high redshifts (Vignali et al. 1999), where most of the
XRB is produced. If anything, this assumption tends to reduce the need
for additional hard spectrum sources, thus strengthening our results.

\begin{figure}
\epsfig{file=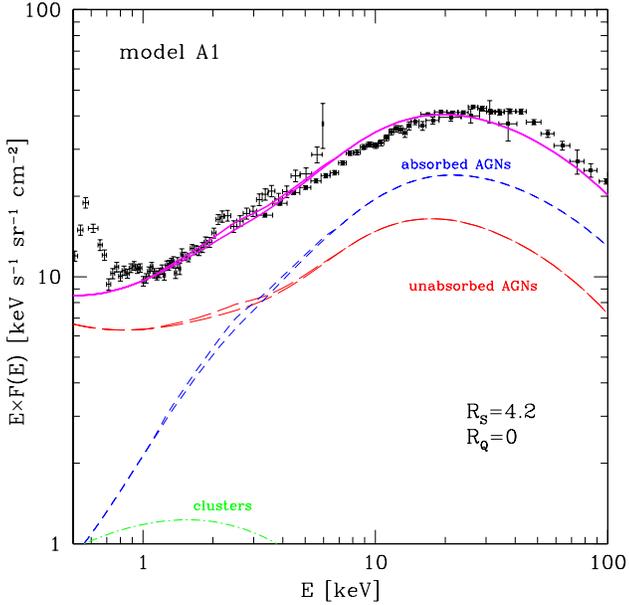,  width=9.cm, height=9.cm,
angle=0}
\caption{The fit to the XRB. The data of higher intensity below 6 keV are
from ASCA (Gendreau et al. 1995), while those above 3 keV are a compendium
of various experiments including HEAO--1 A2 (Gruber 1992). The solid
line is the overall fit while the other labeled curves
represent the contributions of the different classes of sources.
The radiation excess due to the iron line is also shown.}
\end{figure}

Sey 2 spectra have been computed for different amounts of intrinsic
absorption (log$N_{\rm H}$=21.5, 22.5, 23.5, 24.5, 25.5) to cover all the
observed column densities.
In the Compton thin regime the adopted spectrum is that of Sey 1s
with a photoelectric cut off and a lower amount of disc reflection
($f_d=0.88$, Co95). In this regime the component reflected by the torus 
does not contribute significantly to the observed radiation (5\% at 30 keV
for log$N_{\rm H}$=23.5, inclusive of orientation effects). 
For the sources with log$N_{\rm H}$=25.5 we have adopted a pure reflection
continuum. The normalization of the spectrum is determined so as to reproduce
the contribution of thick tori to the flux of Sey 1s (55\% at 30 keV)
after correcting for orientation effects (Ghisellini, Haardt \& Matt,
1994). This approach predicts that the
2--10 keV continuum luminosity of completely Compton thick sources
is about 2\% of the typical luminosity of Sey 1s, in agreement with the
results of Maiolino et al. (1998). 
A composite reflected/transmitted spectrum has been considered for
Circinus--like sources with log$N_{\rm H}$=24.5, where the reflected
and transmitted components have been normalized in analogy with the
previous cases. 
\footnote{For log $N_{\rm H}$=24.5 the effects of Compton scattering begin
to be important. We have checked the error introduced by our approximation 
with respect to the MonteCarlo simulations of Matt, Pompilio \& La Franca  
(1999). The
counts remain unaffected, while the model XRB at 30 keV should be 
reduced by 10--15\%, and even more type 2s should be included in order 
to maintain the agreement with the data.}
  
We have modeled the soft excess of Sey 2s with a power law of index 1.3
and a normalization at 1 keV which is 3\% of the primary
de--absorbed power law. In analogy with type 1s, the spectra of QSO 2s 
--if at all present-- are assumed to be identical to those of Sey 2s.
Finally, we have added to the input spectra
an iron emission line at 6.4 keV. Following Gilli et al. (1999) we have
considered lines with different equivalent widths according to the spectral 
absorption, and have not included the iron line in the spectra of QSOs.

\begin{figure}
\epsfig{file=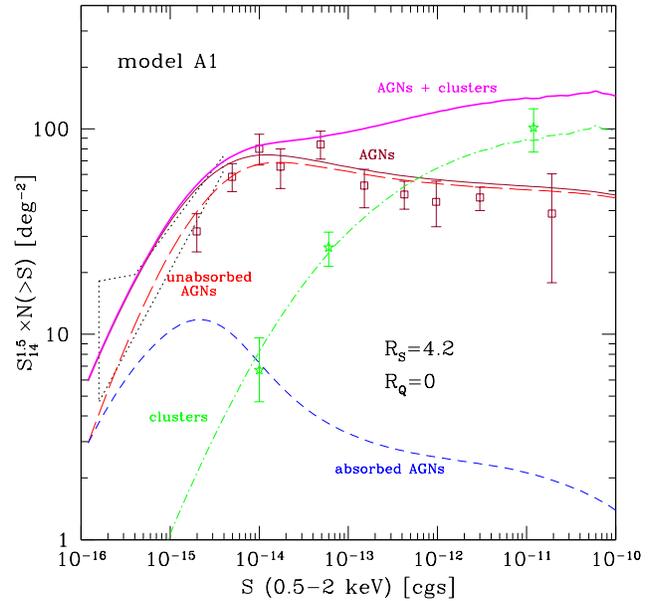,  width=9.cm, height=9.cm, angle=0}
\caption{The predictions of model A1 compared with the
source counts observed in the 0.5--2 keV band by ROSAT. In this and in the
following Figures, source counts are plotted as $S_{14}^{1.5}\times       
N(>S)$, where $S_{14}$ is the flux in units of $10^{-14}$ erg s$^{-1}$
cm$^{-2}$. The AGN data (squares) are adapted from Mi99a.
Cluster counts (stars) at $1\times 10^{-14}$ erg s$^{-1}$ cm$^{-2}$
and $6\times 10^{-14}$ erg s$^{-1}$ cm$^{-2}$ are from
Rosati et al. (1995) and Jones et al. (1998), respectively. The cluster   
surface density at $1.2\times 10^{-11}$ erg s$^{-1}$ cm$^{-2}$ is adapted
from De Grandi et al. (1999).
The fluctuation limits to the total counts (dotted area) are
adapted from Hasinger (1998).}
\end{figure}

In our model A1 we assume $R_{\rm Q}=0$, i.e. we do not include QSO~2s.
The Mi99a
XLF are integrated in the range $10^{41}<L_{\rm x}<10^{49}$ erg s$^{-1}$
up to $z_{max}=4.6$.
The contribution of 
clusters of galaxies has been included by considering thermal
bremsstrahlung spectra with
a distribution of temperatures. We have adopted the 2--10 keV luminosity vs
temperature relation of David et al. (1993), and the 2--10 keV 
X--ray luminosity function of Ebeling et al. (1997). The
cluster XLF is assumed not to evolve, and is integrated in the range
$10^{42}<L_{\rm x}<10^{47}$ erg s$^{-1}$ up to $z_{max}=2$. 

The overall XRB spectrum resulting from the model is shown in Fig.~2 as a
solid line, which is the sum of the contibutions of the other labeled curves.
In order to fit the observed XRB spectrum we need a ratio $R_{\rm S}=4.2$, 
in good agreement with the local value.
Above $\sim 1$ keV, where the XRB is completely extragalactic,
the model provides a good fit to the data from
ASCA (Gendreau et al. 1995) and the compilation of Gruber (1992) based on
HEAO--1 A2 measurements.
The contribution of the AGN iron line to the model XRB is found to be less
than 7\% at $\sim 6.4/(1+z_{cut})$ in agreement with the results of Gilli
et al. (1999) obtained in a different framework (PLE).
Clusters of galaxies are found to contribute to
the model XRB by $\sim 12\%$ at 1 keV, in agreement with the results of
Oukbir, Bartlett \& Blanchard  (1997). 

\subsection{The X--ray source counts}

We now compare the predictions of model A1 with the observed 
source counts in different X--ray bands. The results in the soft 0.5--2
keV band are shown in Fig.~3. 
The expected AGN counts, which are dominated by
unabsorbed sources, agree with the data of Mi99a; the expected cluster
counts agree with the data of Jones et al. (1998), Rosati et al.
(1995), and De Grandi et al. (1999). Since the XLF and its evolution are
derived from the ROSAT counts,
this is no more than a self--consistency check; the slight overprediction
at low fluxes ($\sim$30\% at $\sim 2\times 10^{-15}$ erg cm$^{-2}$ s$^{-1}$) 
is due to the $K$--displaced type~2 objects, as anticipated previously. 

\begin{figure}
\epsfig{file=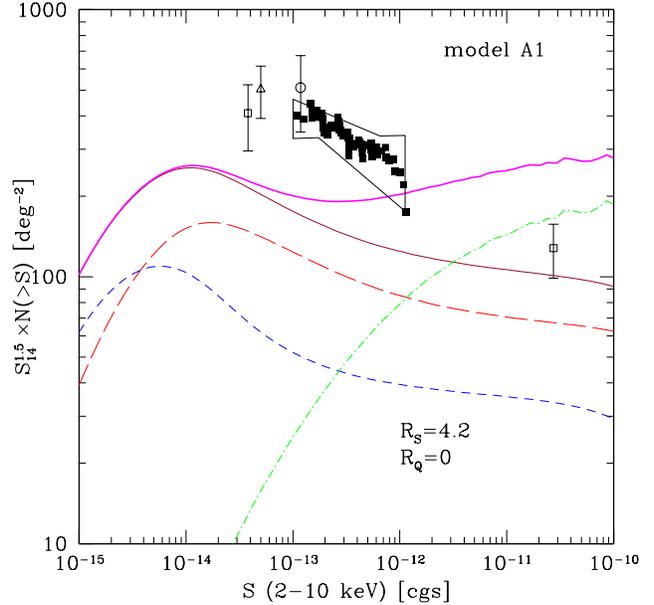, width=9.cm,
height=9.cm, angle=0}
\caption{The predictions of model A1 (lines are as in Fig.~3) compared with the
hard counts between 2 and 10 keV. The open square at $\sim 3\times
10^{-11}$ erg s$^{-1}$ cm$^{-2}$ represents the AGN surface
density in the Piccinotti et al. (1982) sample corrected by 20\% after Co95.
ASCA data are represented by the filled squares (Cagnoni et al. 1998), the
open circle at $\sim 2\times 10^{-13}$ erg s$^{-1}$ cm$^{-2}$
(Ueda et al. 1998) and the open square at $\sim 4\times 10^{-14}$ erg
s$^{-1}$ cm$^{-2}$ (Ogasaka et al. 1998). The open triangle at
$5\times 10^{-14}$ erg s$^{-1}$ cm$^{-2}$ is from BeppoSAX (Giommi et al.
1998).}
\end{figure}

In the hard 2--10 keV band the predictions of the model have to be compared
with the results of HEAO--1 A2 (Piccinotti et al. 1982), ASCA (Cagnoni,
Della Ceca \& Maccacaro 1998; Ogasaka et al. 1998; Ueda et al. 1998)
and BeppoSAX (Giommi et al. 1998).  
At the flux limit of $S\simeq 3\times 10^{-11}$ erg s$^{-1}$ cm$^{-2}$
Piccinotti et al. (1982) found that AGNs and clusters of galaxies have the
same surface density of $1.1\times 10^{-3}$ deg$^{-2}$. However, the AGN
density found by these authors is likely to be overestimated by $\sim
20\%$ due to the local supercluster (Co95). As shown in Fig.~4, after
the Piccinotti et al. point is
corrected by 20\%, the model is in agreement with the data within $1\sigma$. 
On the contrary the disagreement cannot be solved at
fainter fluxes. At $S\sim 2\times 10^{-13}$ erg s$^{-1}$ cm$^{-2}$
the AGN surface density expected in our model
is about a factor of 2 lower than the measurement of Cagnoni et al. (1998).
This corresponds to a $\sim 2\sigma$ discrepancy. When comparing the
model with the
data of the ASCA Large Sky Survey (Ueda et al. 1998) the discrepancy is
even larger. 
  
The situation is worse still in the 5--10 keV band. The only
available counts in this band are from the HELLAS survey
performed by BeppoSAX (Giommi et al. 1998; Comastri et al. 1999). At the 
flux of $\sim 2\times 10^{-13}$ erg s$^{-1}$ cm$^{-2}$ the observed surface 
density is $2.7\pm0.7$ deg$^{-2}$, which is a factor of 4 ($\sim 3\sigma$) 
above the predictions (Fig.~5). 

\begin{figure}
\epsfig{file=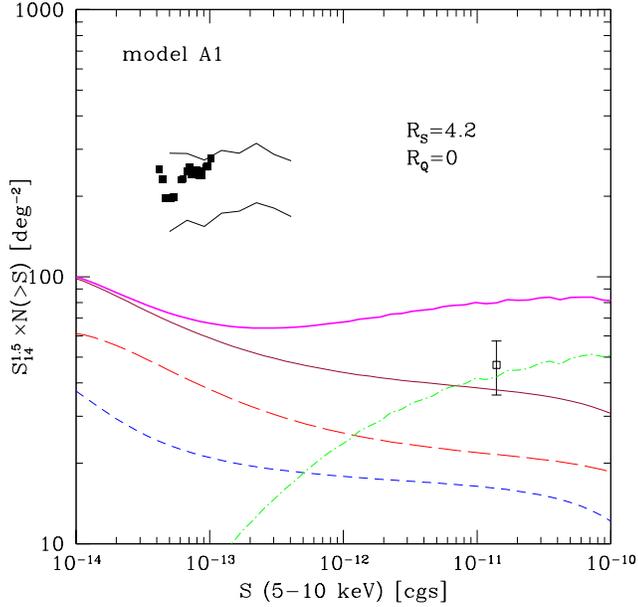,  width=9.cm,
height=9.cm, angle=0}
\caption{The expected counts in model A1 (lines are as in Fig.~3) compared 
with the 5--10 keV BeppoSAX data (Giommi et al. 1998; Comastri et al. 1999).
The open square represents the 20\% corrected AGN surface density found by 
Piccinotti et al. (1982), converted into the 5--10 keV band. The
conversion has been performed by assuming a mean of the input spectra of
our model, weighted for the $N_{\rm H}$ distribution of the Piccinotti et
al. sample (Schartel et al. 1997).}
\end{figure}
  
\subsection{Correction for absorbed sources}

One of the main assumptions of model A1 is that the XLF and 
evolution derived by Mi99a refer to unabsorbed AGNs. However, as discussed
in Sect. 2.2, this is likely not the case. In order to evaluate the
effects of our assumption, we have computed a different variant, A2,
which adopts the XLF and LDDE of Mi99b. These authors allow
self--consistently within their model for the $K$--correction, and for
the absorbed sources which, especially at faint fluxes and high redshifts,
appear in the ROSAT counts; thus the parameters they provide refer
to unabsorbed sources only. Of course, our model is different from theirs,
and the self--consistency is lost; however, this is likely to be a higher
order effect, and for a first order estimate we can include their
parameters in our computation. The results are shown in Fig.~6. Note that
around 1 keV the unabsorbed sources produce 30\% of the XRB, exactly as in
Mi99b. Note also that in
order to fit the XRB spectrum a ratio $R_{\rm S}=13$ is now required, which is 
much higher than the local value and implies additional hard spectrum sources. 
Since the contribution of type 2s has increased with respect to A1, in order 
to make up for the reduced contribution of type 1s, the mean spectrum of the 
population producing the XRB is harder, and the discrepancies between the
model predictions and the hard counts are somewhat reduced (though not
completely eliminated). 

\begin{figure} 
\epsfig{file=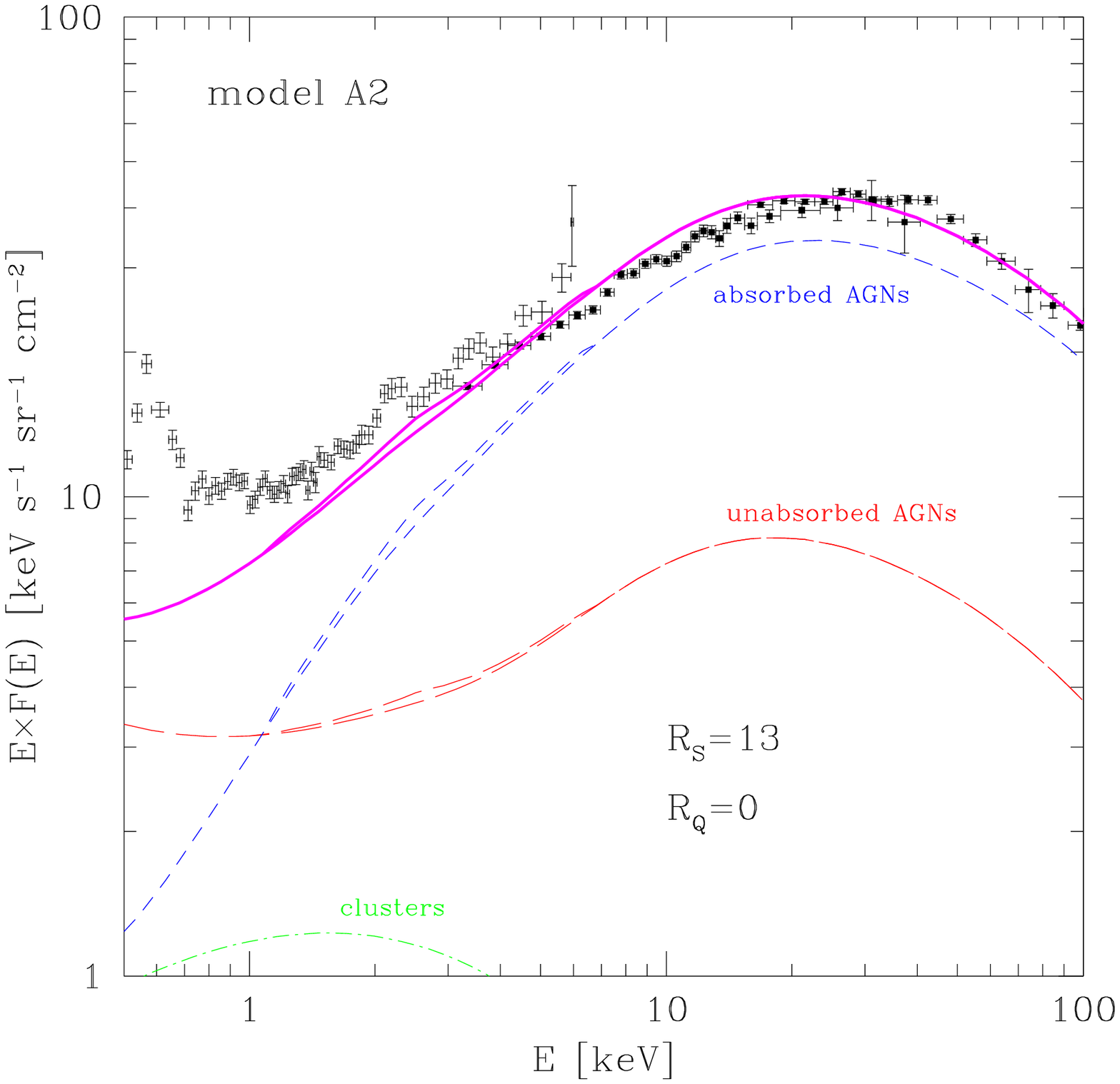,  width=9.cm, height=7.cm,
angle=0}
\epsfig{file=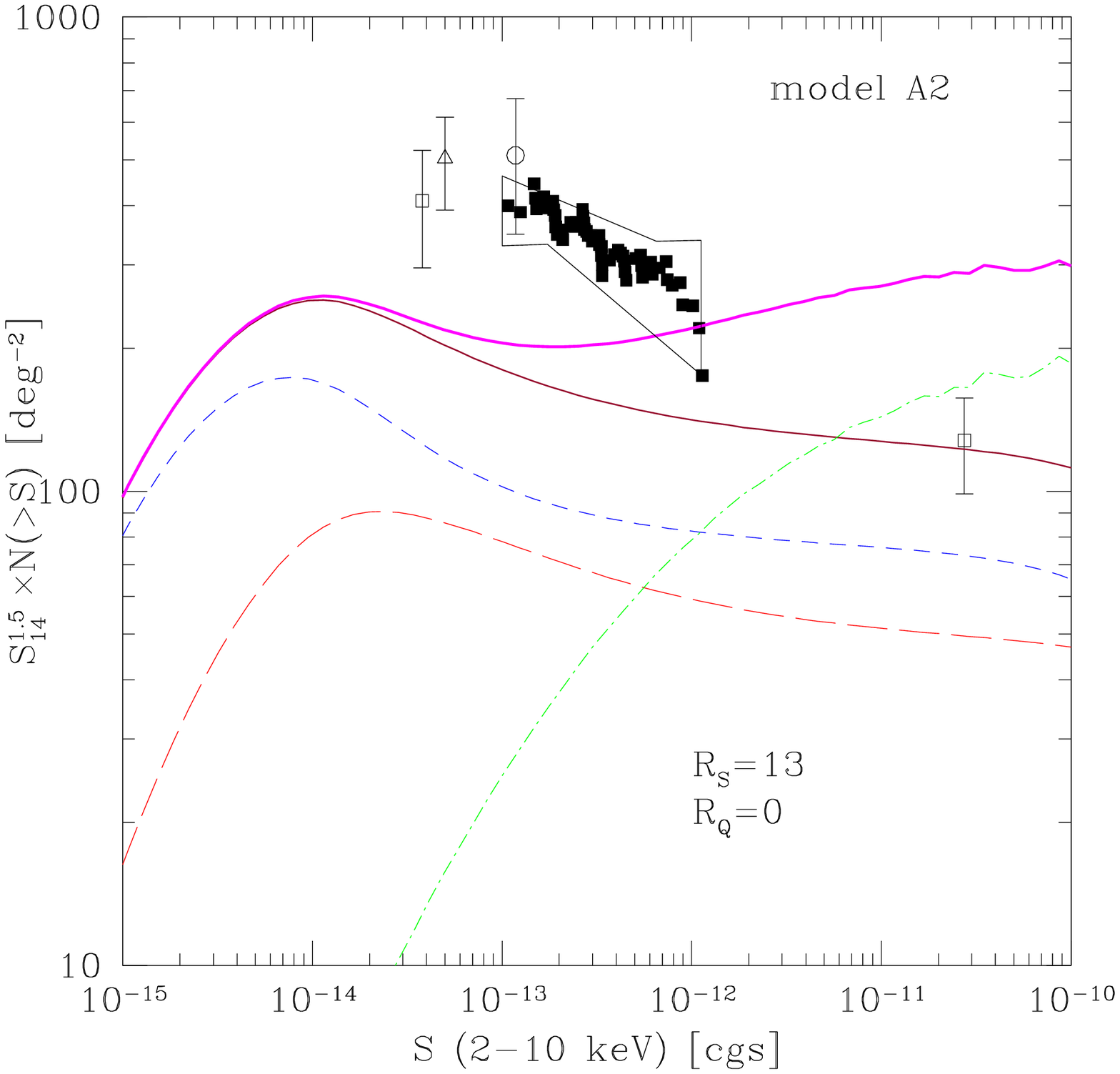,  width=9.cm, height=7.cm,
angle=0}
\epsfig{file=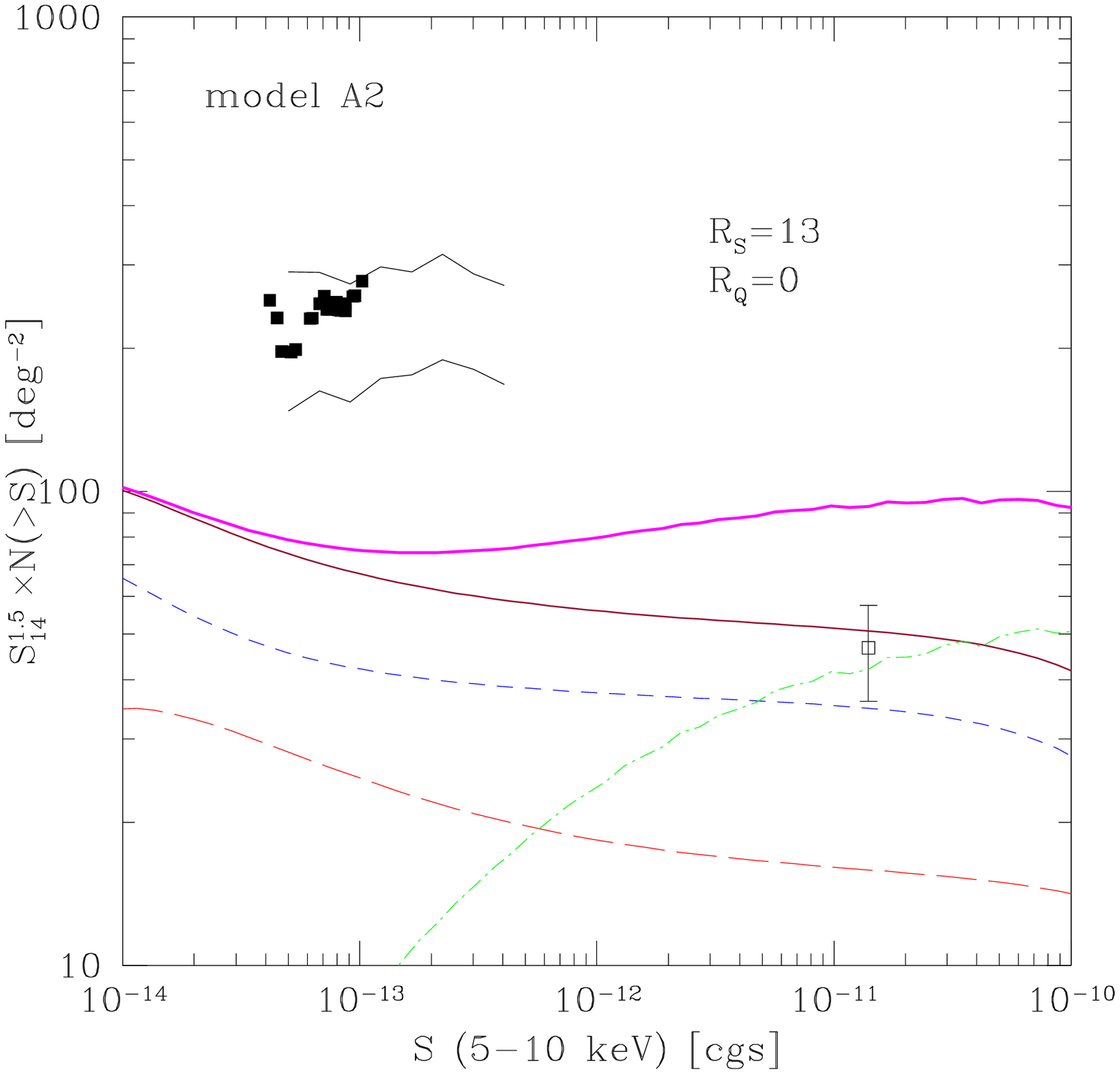,  width=9.cm, height=7.cm,
angle=0}
\caption{From top to bottom, the same as in Figs.~2, 4, 5 but for model 
A2. Curves are as in the previous Figures.
Note the very hard spectrum of the model XRB due
to the high contribution of absorbed sources.}
\end{figure}

\section{Discussion}

The main difference between models A1 and A2 is the fractional contribution
of type 1 AGNs to the XRB; this contribution is dominated by objects close
to the XLF break at redshifts close to $z_{cut}$, and is not well constrained
by the data. In the former model 60\% of the 1--keV XRB is due to type 1s, so
the local value of the type ratio $R_{\rm S}$ is sufficient to account for the
entire XRB; the average spectrum, though, is too soft, and the softness shows
up in a marginal discrepancy with the XRB spectrum at $>40$~keV (Fig.~2),
and unacceptable discrepancies with the hard counts (Figs.~4 and
5). In the latter model the type 1s account for only 30\% of the 1--keV XRB, 
and making up the entire XRB requires an $R_{\rm S}$ much larger than the local
value; now the average spectrum is harder, the shape discrepancy disappears
and the count discrepancies are reduced (Fig. 6). By extrapolating from 
these two models we can make qualitative predictions on still different 
parametrizations of the Mi99a sample: for instance, models adopting density
evolution with a dependence on luminosity weaker than Mi99a would predict a
type 1 soft X--ray contribution higher than 60\%, and would miss the hard 
counts by factors larger than Figs.~4 and 5. Density evolution with a
dependence on luminosity stronger than Mi99b (if at all acceptable) would
require $R_{\rm S}>13$, which is already three times the local value.

The main result of our analysis is precisely this one: no matter which
variant is adopted for the XLF and the evolution, the models which
incorporate the most recent observations within the standard prescriptions 
always produce some discrepancy. The discrepancy may appear as an 
underprediction of the observed hard counts, or a type 2 to type 1 ratio
higher than the observed local value, but in all cases it points to
additional hard spectrum sources at intermediate or high redshifts.

For the sake of completeness, we present in the Appendix a PLE model 
with $R_{\rm Q}=R_{\rm S}$ (model B): this region of the parameter space
is not favored by the most recent data, but was adopted in practically
all previous works on the XRB. Note that model B has the same type 1 soft 
X--ray contribution as model A2 (30\%), but the type 2 contribution here
is due to higher luminosity sources, which show up in 
higher flux bins: indeed, the XRB spectrum is well fitted, the counts in 
the ASCA band are matched, and the discrepancy with the HELLAS counts is
reduced to $2\sigma$. The cost to be paid is a number density 
$R_{\rm Q}=7.7$, which --again-- is definitely higher than the local upper 
limit $R_{\rm Q}<2$. Irrespective of the plausibility of PLE and QSO~2s,
we stress that even model B results in a discrepancy, and the discrepancy
is concordant with the results of the other, less controversial variants.

A population of absorbed or hard spectrum AGNs evolving more rapidly than
the type~1s could accomodate all the problems discussed above. In this
context one should be reminded that the hard counts already resolve
$\sim$30\% of the XRB at fluxes $\sim~5\times 10^{-14}$ erg cm$^{-2}$
s$^{-1}$, so they must converge rapidly just below these values. 
The optical identifications of the counts in the 2--10 keV and
5--10 keV bands are still largely incomplete.
Up to now 34 X--ray sources detected in the ASCA LSS survey (Ueda et al.
1998) have been identified (Akiyama et al. 1998), and 28 objects turned out
to be AGNs. They are 22 broad line AGNs (type 1--1.5) with
$0<z<1.7$, and 6 type 2 AGNs with $0<z<0.7$. The number of identified
sources of the BeppoSAX HELLAS survey is lower, but the distribution of
the AGNs seems similar to the ASCA one: 7 broad
line QSOs with $0.2<z<1.3$ and 5 Seyferts 1.8--1.9 with $0.04<z<0.34$
(Fiore et al. 1999).

If one accepts these low redshift type 2 identifications, 
one has to find a physical reason for a 
convergence so recent in comparison with all other AGNs (BL--Lacs excepted) 
and star forming galaxies. Alternatively, one could rely on the poor
statistics 
to maintain that the hard counts are mostly due to optically empty fields, 
containing very absorbed, very powerful sources at redshifts $>1$.

There are prospective candidates in both scenarios. In the low--$z$
hypothesis one could assume that ``normal'' Seyfert 2 galaxies evolve 
more rapidly than type 1s, so that $R_{\rm S}$ increases with redshift
up to the required value. Not only the number ratio, but also the $N_{
\rm H}$ distribution could change with cosmic time (Franceschini et al. 1993). 
Local Sey 2s are
associated with a star formation activity higher than Sey 1 and normal
galaxies ( Maiolino et al. 1995; Rodr\'{\i}guez-Espinoza et al. 1986),
so this assumption would have interesting implications on the star
formation history. One could also
invoke Advection Dominated Accretion Flows (ADAFs,
Di Matteo et al. 1998) whose luminosity is proportional to $\dot M^2$,
where $\dot M$ is the mass accretion rate,
and which become normal QSOs at large $\dot M$: thus, they should
evolve more rapidly than normal AGNs at intermediate redshifts, and
should undergo a change of class at high redshifts. In the high--$z$
hypothesis one should resort to ULIRGs, which indeed are absorbed and
powerful, and appear to evolve as fast as required [$(1+z)^{7.6\pm3.2}$,
Kim \& Sanders 1998]. As mentioned before there is evidence that the IR 
emission of ULIRGs is powered both by starburst and AGN processes; 
Kim, Veilleux \& Sanders (1998) and Lutz et al. (1998) find that the
fraction
of AGN--powered infrared luminous galaxies increases with the bolometric
luminosity, and reaches 30--50\% in the ULIRG range. While normal starbursts 
are inefficient emitters in the hard X--rays, obscured AGNs in ULIRGs 
could easily explain the hard counts.

Finally, it should be noted that the optical identifications of the hard
X--ray counts, scanty as they are, suggest that at the given X--ray flux
the type 1s are more numerous than the type 2s. Concordant evidence is
provided by recent BeppoSAX observations of the Marano field and the
Lockman hole (Hasinger et al. 1999): most of the BeppoSAX sources have
ROSAT counterparts, which in most cases are optically identified with
type 1 AGNs. This type composition is in agreement with the predictions
of, for instance, model A1 (Figs.~4 and 5). But if the hard counts in
excess of the model are attributed entirely to obscured AGNs, then the
predicted type ratio is reversed, with the type 2s more numerous than the
type 1s. The numbers involved are too small to draw any conclusion,
however they seem to suggest that some of the hard counts are due to flat
X--ray spectrum sources with type 1 optical spectra; a few similar sources
might have been found already in the ASCA LSS (Akiyama et al. 1998).

Clearly, a decisive progress in this area will require more numerous
and more secure identifications of hard X--ray counts; given the various
hypotheses, counterparts should be looked for not only at optical
wavelengths, but also in the infrared and submillimeter domains, where
AGN--dominated ULIRGs should be conspicuous. 

\section{Summary and conclusions}

In this paper we have shown that the standard prescriptions for
synthesizing the XRB from the integrated emission of AGNs are 
not consistent with a number of recent observational constraints,
and some of them must be relaxed.

We have worked out models (A1 and A2) which take into account detailed
input spectra of AGNs, the $N_{\rm H}$ distribution observed in local
Seyfert 2s, and the XLF and evolution newly determined from the largest
ROSAT sample. The latter data do not define a unique parametrization, 
and the two models explore different variants. As prescribed by the
standard model, the XLF and evolution of type 2 AGNs are taken from
type 1s, and the spectra of both types are taken independent of redshift;
the only fitting parameter is the number ratio $R$ of type 2s to type 1s.
We find that model A1 reproduces the XRB and the soft counts with a ratio
$R$ compatible with the local value, but underestimates the hard counts. 
Model A2 is less discrepant as far as the counts are concerned, but requires 
a ratio $R$ definitely larger than observed locally.
We have also computed a model adopting a canonical pure luminosity 
evolution (model B).
In agreement with the results of Co95, model B can reproduce the 
XRB, the soft X--ray counts and the ASCA hard counts in the
2--10 keV band. It is also consistent within 2$\sigma$ (or discrepant at
2$\sigma$) with the preliminary BeppoSAX counts in the 5--10 keV band.
Nevertheless, it requires a number of type 2 QSOs much higher than the local
upper limit, and perhaps already ruled out by the deep X--ray surveys.

The discrepancies found in all models are to some extent model dependent,
but all of them point in the same direction, and suggest that hard 
spectrum
sources at intermediate or high redshifts are needed in addition to the
predictions of the standard scenario.
The X--ray spectrum of these additional sources could be flattened by
absorption,
or could be intrinsically hard. In the former hypothesis 
reasonable candidate counterparts could be rapidly evolving, ``normal''
Seyfert 2s. One should also note that a fraction of ULIRGs
seem to be powered by AGNs, and their cosmological evolution 
seems faster than that of unabsorbed QSOs.
The alternate hypothesis could instead require the presence of ADAFs.  
Optical identifications of the hard X--ray sources are still largely
incomplete and do not allow yet to decide between the various possibilities.

\begin{acknowledgements}

We are grateful to A. Comastri and G. Zamorani for a careful reading of
the manuscript, and to T. Miyaji G. Hasinger and M. Schmidt for permission
to use their LDDE model in advance of publication. Our presentation was
greatly improved by the comments of the referee, Prof. G. Hasinger. This 
work was partly supported by the Italian Space Agency
(ASI) under grant ARS--98--116/22 and by the Italian Ministry for
University and Research (MURST) under grant Cofin98--02--32. 

\end{acknowledgements}

\appendix
\section{Comparison with a PLE model}
\begin{figure}
\epsfig{file=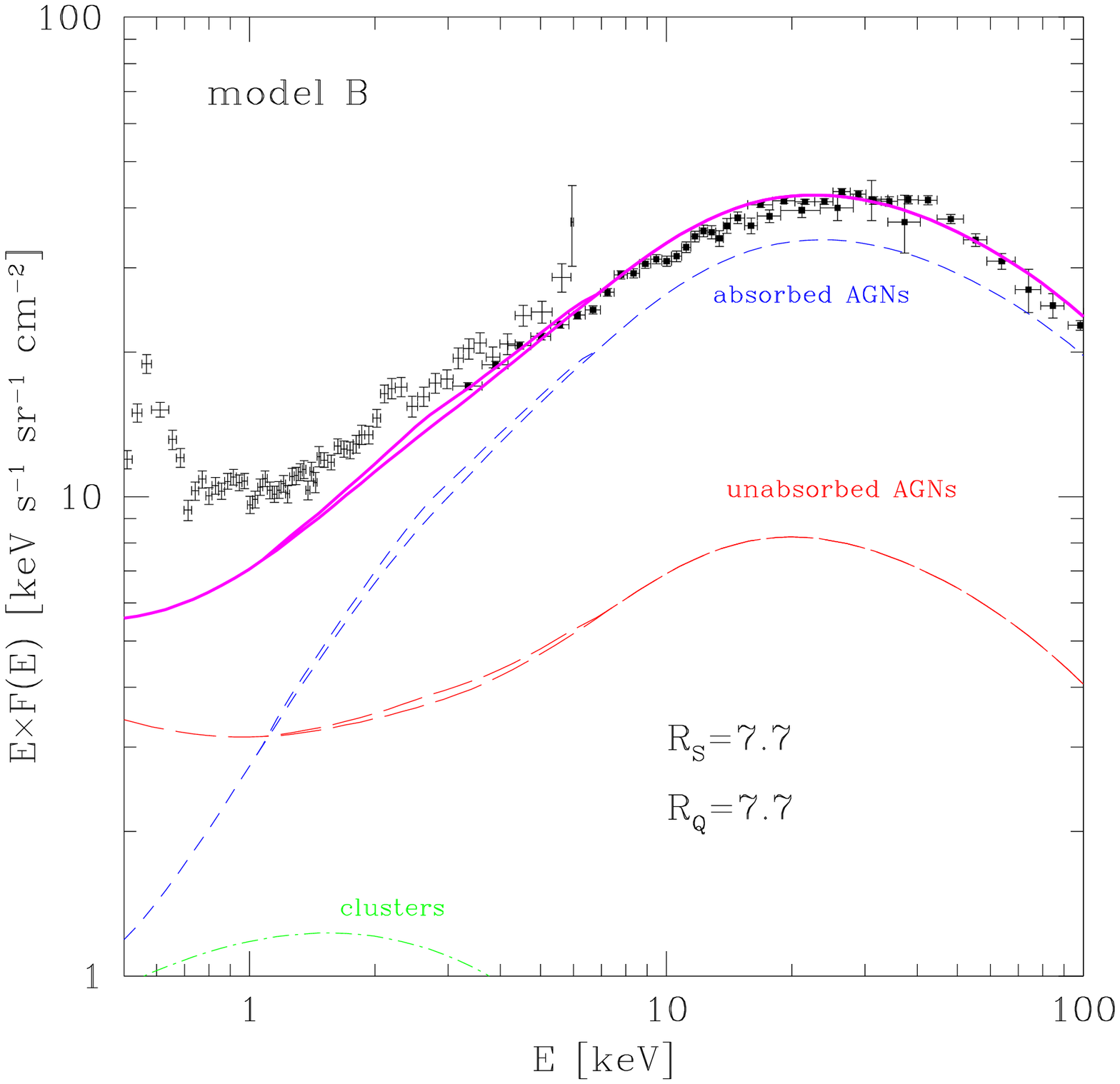,  width=9.cm, height=7.cm,
angle=0}
\epsfig{file=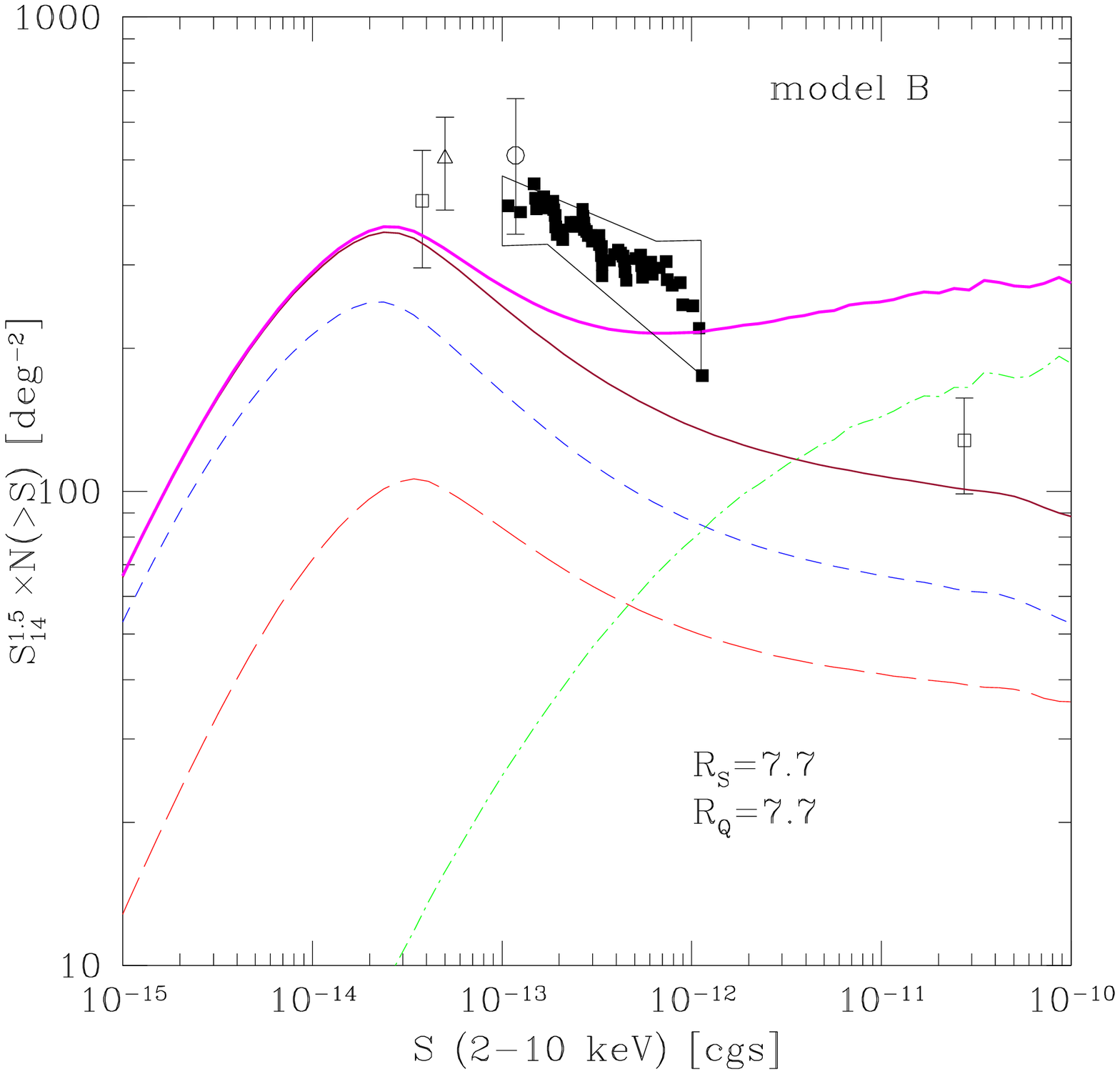,  width=9.cm, height=7.cm,
angle=0}
\epsfig{file=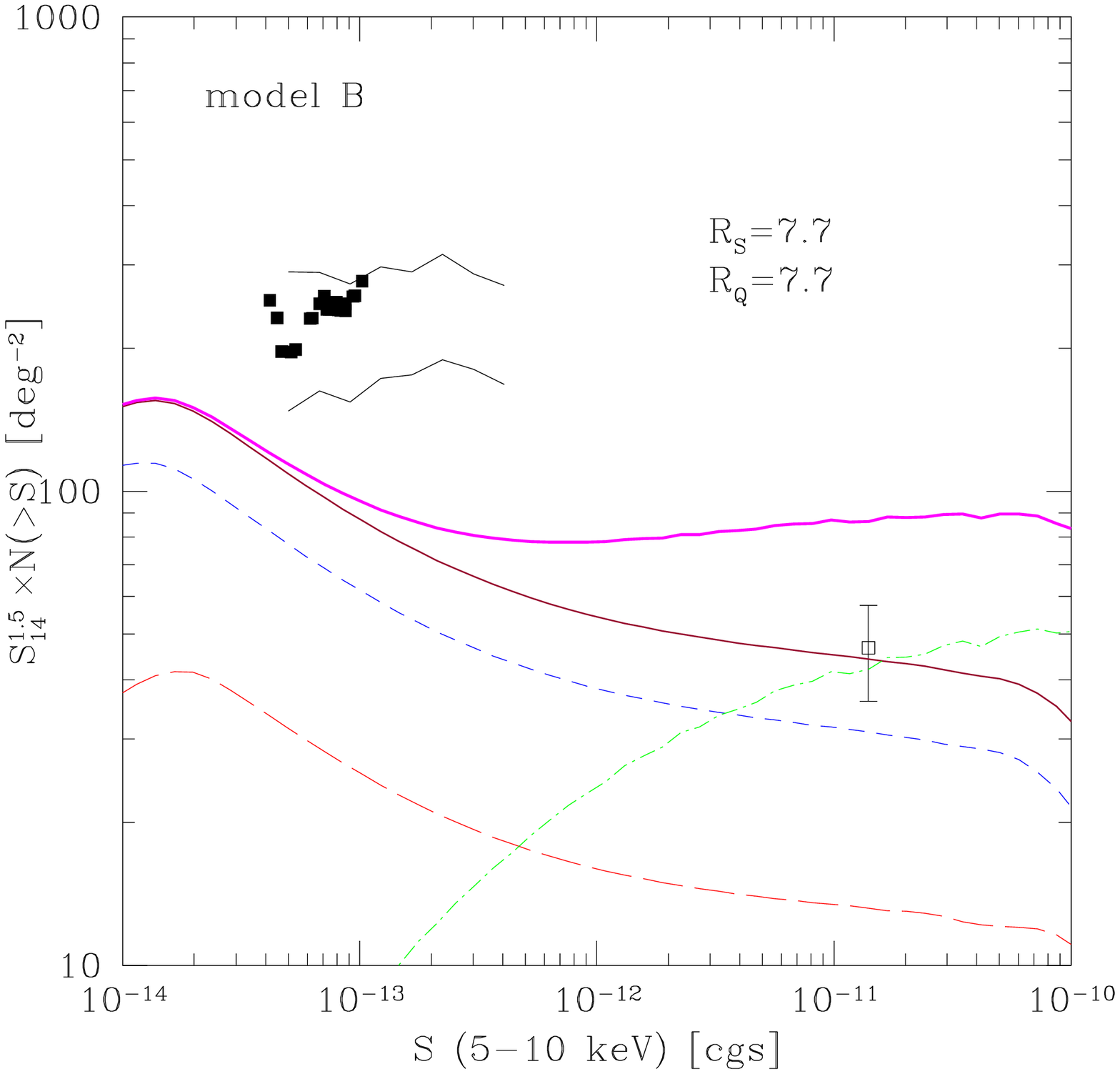,  width=9.cm, height=7.cm,
angle=0}
\caption{From top to bottom the same as in Figs. 2, 4, 5, but for model
B.}
\end{figure}

We have computed a canonical synthesis model of the XRB
by adopting the XLF and PLE of Jones et al. (1997); since only AGNs with
broad optical lines are included, there is no need to correct for the
contribution of type 2 AGNs. This sample is smaller than Mi99a, and
presumably low luminosity sources at high redshifts are
under\-re\-pre\-sented.
Our model B assumes the XLF and PLE indicated above, includes QSO 2s
as numerous as the Sey 2s, and adopts the absorption distribution
of Risaliti et al. (1999) at all redshifts and all luminosities.
In the cosmology adopted here, model B makes only $\sim 30\%$ of the soft 
XRB with type 1 AGNs, and one needs $R_{\rm S}=R_{\rm Q}=7.7$ to fit the
overall background (Fig.~A1). Due to the large contribution
of type 2 AGNs, the model XRB spectrum is very hard.
Furthermore, due to the high ``effective'' luminosity implied by QSO~2s,
the ASCA counts are reproduced. The discrepancy with the data in
the 5--10 keV band, on the contrary, is not eliminated, although it is
reduced to a $2\sigma$ level. Because of the preliminary nature
of the HELLAS data one might debate about its significance. At any rate,
one
should stress that this marginal result can be obtained only by assuming a
strong (a factor $>$4) differential evolution of QSO 2s with respect to
QSO 1s, so that at $z_{cut}$ the former would outnumber the latter by
a factor $\sim$8.

\end{document}